\documentstyle[epsf,epsbox]{article}
\begin{document}
\vspace{ .7cm}
\vspace*{2cm}
\begin{center}
{\LARGE\bf CP Violation in the semileptonic top decay in two-Higgs doublet model}\\
\vspace{1 cm}
{\large Tsutom \ Hasuike$ ^{a), }$\footnote{e-mail: hasuike@anan-nct.ac.jp},  Toshihiko \ Hattori$ ^{b), }$\footnote{ e-mail: hattori@ias.tokushima-u.ac.jp}, Seiichi \ Wakaizumi$ ^{c), }$\footnote{e-mail: wakaizum@medsci.tokushima-u.ac.jp}}

\vspace {1cm}
a) Department of Physics,\ Anan College of Technology, 774-0017, Japan \\
b) Institute of Theoretical Physics, University of Tokushima, 770-8502, Japan \\
c) School of Medical Science, University of Tokushima, Tokushima 770-8509, Japan
\end{center}
\vspace{2cm}
{\Large\bf Abstract}\\  
CP violation in semileptonic top-quark decay is investigated by exactly using one charged- and three neutral-Higgs bosons obtained by solving the Higgs mass matrix in two-Higgs doublet model. The CP-violating up-down asymmetry of leptons from $W$ boson decays is shown to be $1 \times 10^{-4} \sim 4 \times 10^{-3}$ for the region of $\tan \beta \ll 1$, where $\tan \beta$ is the ratio of vacuum expectation values for the two neutral Higgs bosons.
\newpage
\setcounter{footnote}{0}
\def\thefootnote{\arabic{footnote}}

\section{ Introduction}

CP violation in kaon physics was explained by the phenomenological ${\rm K}^0 - \overline{\rm K^0}$ mixing.
It has been suggested that CP violation in this sysytem originates in the Kobayashi-Maskawa mechanism of CP violation \cite{KM}. Its mechanism is expected to be tested by B-meson decays in the coming  B-factories. CP violation in top-quark decays will also be studied in the future experiments. 

The top-quark decays have some advantages superior to the K- and B-decays, because the top quark would decay before it  hadronizes and we can use its polarization in order to detect CP violation \cite{KLY}, since the lifetime is shorter than $10^{-23}$ s \cite{BDKKZ} due to its large mass ($m_t = 180$ GeV) \cite{CDF}. CP violation in $t \rightarrow W^+ b$ decays is estimated to be very small in the Kobayashi-Maskawa Standard Model and therefore it is sensitive to Higgs boson exchanges of the non-standard effect. The two-Higgs doublet model of type II \cite{GHKD} with the CP-violating neutral sector \cite{BR} was used in the study of electric dipole moment of neutron \cite{W}, and then the idea was applied to $t \rightarrow W^+ b$ decay \cite{GG}. Lately, we considered consistently in unitary guage the CP violating up-down asymmetry of the leptons from W boson decays in $t \rightarrow W^+ b$ \cite{HHW}, which was defined by Grz{\c a}dkowski and Gunion \cite{GG}. Though only one!
 degenerate neutral Higgs boson was used for calculating the effects of neutral Higgs boson exchange in those papers, the exact number of the real neutral Higgs bosons is three in the two-Higgs doublet model \cite{HKMT}. In the present paper, by using one charged- and three neutral-Higgs bosons obtained by solving the Higgs mass matrix, we calculate the CP-violating up-down asymmetry in the top-quark decay. In the next section, the asymmetry of lepton distributions is briefly expressed by the non-standard $tbW$-vertex form factors. In Sect. 3 Higgs mass matrix is investigated in the two-Higgs doublet model, and then the CP-violating asymmetry is examined exactly with the three neutral Higgs bosons in Sect. 4. Section 5 is devoted to the conclusions and discussions.

\section{Up-down asymmetry of lepton distributions in top decays}

We investigate CP violation in $t \rightarrow W^+b$ decays \cite{HHW} by the lepton distribution up-down asymmetry from W decay defined by Grz{\c a}dkowski and Gunion \cite{GG}. We introduce the up-down asymmetry with respect to $t \rightarrow W^+ b$ decay plane in the top-quark laboratory frame for $e^+e^- \rightarrow t{\bar t}$ defined in Ref. 8 as
\begin{equation}
 A^t \equiv \frac{N^t}{D^t} \ ,
\end{equation} 
where
\begin{equation}
N^t = \int {\rm d}\Phi (t \rightarrow W^+ b) \int_1^{-1} {\rm d}\cos \theta_{l^+} \left[ \int _0^\pi  - \int_{-\pi}^0 \right] {\rm d}\phi_ {l^+} \frac{{\rm d}\sigma_{\rm tot}}{{\rm d}\Phi (t \rightarrow W^+ b) {\rm d} \Phi (W^+ \rightarrow l^+ \nu)},
\end{equation}
\begin{equation}
D^t = \int {\rm d}\Phi (t \rightarrow W^+ b) \int_1^{-1} {\rm d}\cos \theta_{l^+} \left[ \int_0^\pi+ \int_{-\pi}^0 \right] {\rm d}\phi_ {l^+} \frac{{\rm d}\sigma_{\rm tot}}{{\rm d}\Phi (t \rightarrow W^+ b) {\rm d} \Phi ( W^+ \rightarrow l^+ \nu)},
\end{equation}
in which $\theta_{l^+}$ and $\phi_{l^+}$ are the polar and azimuthal angles of the lepton $l^+$ in the $W^+$ rest frame, and ${\rm d}\sigma_{\rm tot}/\lbrack{\rm d}\Phi (t \rightarrow W^+ b){\rm d}\Phi (W^+ \rightarrow l^+ \nu)\rbrack$ is the differential cross section for $t \rightarrow W^+ b \rightarrow l^+\nu b$, ${\rm d}\Phi (t \rightarrow W^+ b)$ being the phase space for $t \rightarrow W^+b$ decay and so on. In calculating Eq. (2) and Eq. (3), we adopt the most general parametrization of $tbW$-vertex for $t \rightarrow W^+ b$ decays as follows,
\begin{equation}
\Gamma^\mu = -\frac{g}{\sqrt{2}} V_{tb}\bar{u}(p_b) \lbrack \gamma^\mu (f^L_1 P_L + f^R_1P_R ) - \frac{i\sigma^{\mu \nu}k_{\nu}}{m_W} (f^L_2P_L + f^R_2P_R ) \rbrack u(p_t), 
\end{equation} 
where $P_{R/L} =(1 \pm \gamma_5)/2$, $k$ is the $W$ momentum, $V_{tb}$ is the $(tb)$-element of the Cabibbo-Kobayashi-Maskawa mixing matrix and $g$ is the SU(2) guage coupling constant. We obtain the following expression for $N^t$ and $D^t$ as developed in Refs. 8 and 9,
\begin{eqnarray}
N^t &=& A\ 2\pi^3\frac{m_t^2 - m_W^2}{m_W^2} P_{||}^t {\rm Im}(f_1^L f_2^{R*} ), \\
D^t &=& A \frac{16\pi^2}{3}\lbrack \frac{m_t^2 + 2m_W^2}{m_W^2} \vert f_1^L \vert^2 + 6\frac{m_t}{m_W} {\rm Re}(f_1^Lf_2^{R*}) + \frac{2m_t^2 + m_W^2}{m_W^2}\vert f_2^R\vert^2\rbrack,
\end{eqnarray}
where $A$ is a common factor and $P_{||}^t$ is the longitudinal polarization of top quark. If we keep only leading term in $D^t$, we obtain the following expression for the asymmetry $A^t$,
\begin{equation}
A^t = \frac{3\pi}{8} \frac{m_t^2 - m_W^2}{m_t^2 + 2m_W^2} P_{||}^t {\rm Im}(f_1^Lf_2^{R*})/{\vert f_1^L\vert}^2.
\end{equation}
It is important to note that $V_{tb}$ disapears in $A^t$ and $f_1^L = 1$ and $f_2^R = 0$ at the tree level and $f_2^R$ does not have any contribution even at the one-loop level in the Standard Model \cite{GK}. So, $A^t$ is sensitive to the non-standard origin of CP violation.

\section{Higgs mass matrix in two-Higgs doublet model}

The two-Higgs doublet model (type II) with the CP-violating neutral sector was adopted to calculate the neutron electric dipole moment by Weinberg \cite{W}. Though he used the approximation of one degenerate neutral Higgs boson, this degeneracy is dissolved into three Higgs bosons by diagonalizing the Higgs mass matrix \cite{HKMT}. By using one charged- and three neutral-Higgs bosons coming from the two-Higgs doublets $\phi_1$ and $\phi_2$, we reexamine CP violation in top-quark decays. We adopt the parametrization for Higgs bosons defined by Weinberg \cite{W}, which is formulated in unitary guage through the unitary guage condition \cite{W2}, where Goldstone bosons do not appear in this model. The three neutral scalars in the model are parametrized as

\begin{eqnarray}
\phi_1^0 & = & \frac{v_1}{\sqrt{2}\vert v_1\vert}\left[\Phi_1 - i\frac{\vert v_2\vert}{v}\Phi_3\right], \nonumber \\
\phi_2^0 & = & \frac{v_2}{\sqrt{2}\vert v_2\vert}\left[ \Phi_2 - i\frac{\vert v_1\vert}{v} \Phi_3\right],
\end{eqnarray}
where $v_1$ and $v_2$ are the vacuum expectation values of $\phi_1$ and $\phi_2$, respectively and $v \equiv \sqrt{ |v_1|^2 + |v_2|^2}$. The two real new fields $\Phi_1$ and $\Phi_2$ are of CP-even and the third scalar $\Phi_3$ is of CP-odd, as is evident from Eq. (8), so CP violation will occur via the scalar-pseudoscalar interference term in the following quantities,
\begin{eqnarray}
\frac{\langle \phi_2^0\phi_1^{0*}\rangle}{v_1^*v_2} & = & \sum_{k = 1}^3 \frac{\sqrt{2}G_FZ_0^{(k)}}{m_k^2 - q^2}, \quad
\frac{\langle \phi_2^0\phi_1^0\rangle}{v_1v_2}  =  \sum_{k = 1}^3 \frac{\sqrt{2}G_F{\tilde Z}_0^{(k)}}{m_k^2 - q^2}, \nonumber \\
\frac{\langle \phi_1^0\phi_1^0\rangle}{v_1^2} & = & \sum_{k = 1}^3 \frac{\sqrt{2}G_FZ_1^{(k)}}{m_k^2 - q^2}, \quad
\frac{\langle \phi_2^0\phi_2^0\rangle}{v_2^2}  =  \sum_{k = 1}^3 \frac{\sqrt{2}G_FZ_2^{(k)}}{m_k^2 - q^2},
\end{eqnarray}
where $\langle \phi_i^0\phi_j^0\rangle$ is for any pair of scalar fields $\phi_i^0$ and $\phi_j^0$, an abreviation for the momentum-dependent quantity
\begin{equation}
\langle \phi_i^0\phi_j^0\rangle = \int {\rm d}^4x\langle T\lbrack\phi_i^0(x) \phi_j^0 (0)\rangle_{\rm vac} e^{-iq\cdot x},
\end{equation}
evaluated in the tree-level approximation, and $m_k \ (k = 1, 2, 3)$ are the masses of real neutral scalar Higgs bosons, the eigenvalues of the Higgs mass matrix. The imaginary parts of the scalar fields normalization constants $Z_i \ (i = 0,1, 2$) are deduced to 
\begin{eqnarray}
{\rm Im} Z_0^{(k)} & = & \sqrt{1 + \cot^2\beta} \ u_1^{(k)}u_3^{(k)} + \sqrt{1 + \tan^2\beta}\ u_2^{(k)}u_3^{(k)},\nonumber \\
{\rm Im} {\tilde Z}_0^{(k)} & = &  \sqrt{1 + \cot^2\beta}\ u_1^{(k)}u_3^{(k)} - \sqrt{1 + \tan^2\beta}\ u_2^{(k)}u_3^{(k)},\nonumber \\
{\rm Im} Z_1^{(k)} & = & -2\sqrt{\tan^2\beta + \tan^4\beta}\ u_1^{(k)}u_3^{(k)}, \nonumber \\
{\rm Im} Z_2^{(k)} & = & 2\sqrt{\cot^2\beta + \cot^4\beta}\ u_2^{(k)}u_3^{(k)},
\end{eqnarray}
where $u_i^{(k)}$ denotes the $i$-th component of the $k$-th normalized eigenvector of the Higgs mass matrix and $\tan \beta = \vert v_2\vert /\vert v_1\vert$. 

Let us estimate $u_i^{(k)}$ and $m_k$ by studying the symmetric Higgs mass matrix squared $M^2$ derived from the Higgs potential
\begin{eqnarray}
V &=& \frac{1}{2}g_1(\phi_1^\dagger \phi_1 - {\vert v_1 \vert}^2)^2 + \frac{1}{2} g_2 (\phi_2^\dagger\phi_2 - {\vert v_2 \vert}^2)^2  \nonumber \\
&& \mbox{} +g_3 (\phi_1^\dagger \phi_1 - {\vert v_1\vert}^2)(\phi_2^\dagger\phi_2 - {\vert v_2\vert}^2) +g'{\vert\phi_1^\dagger\phi_2 - v_1^*v_2\vert}^2 \nonumber \\
&& \mbox{} + {\rm Re}\lbrack h(\phi_1^\dagger\phi_2 - v_1^*v_2)^2\rbrack + \xi \lbrack \frac{\phi_1}{v_1} - \frac{\phi_2}{v_2}\rbrack^\dagger\lbrack\frac{\phi_1}{v_1} - \frac{\phi_2}{v_2}\rbrack.
\end{eqnarray}
We insert Eq. (8) into Eq. (12) and get the symmetric Higgs mass matrix $M^2$, whose elements are 
\begin{eqnarray}
M_{11}^2 &=& 2g_1{\vert v_1 \vert}^2 + g'{\vert v_2\vert}^2 + \frac{\xi + {\rm Re}( hv_1^{*2}v_2^2)}{{\vert v_1\vert}^2}, \nonumber \\
M_{22}^2 &=& 2g_2{\vert v_2 \vert}^2 + g'{\vert v_1\vert}^2 + \frac{\xi + {\rm Re}( hv_1^{*2}v_2^2)}{{\vert v_2\vert}^2},\nonumber \\
M_{33}^2 &=& v^2\lbrack g' + \frac{\xi - {\rm Re}( hv_1^{*2}v_2^2)}{{\vert v_1v_2\vert}^2}\rbrack ,\nonumber \\
M_{12}^2 &=& \vert v_1 v_2\vert (2g_3 + g') + \frac{{\rm Re}( hv_1^{*2}v_2^2) - \xi}{\vert v_1v_2\vert},\nonumber \\
M_{13}^2 &=&  -\frac{v}{\vert v_1^2v_2\vert}{\rm Im}(hv_1^{*2}v_2^2),\nonumber \\
M_{23}^2 &=& -\frac{v}{\vert v_1v_2^2\vert} {\rm Im}(hv_1^{*2}v_2^2).
\end{eqnarray}
As a phase convention, we take $h$ to be real and
\begin{equation}
v_1^*v_2 = \vert v_1 \vert \vert v_2 \vert {\rm exp}(i\phi).
\end{equation}
The parameters are constrained only by the following positivity conditions \cite{W2},
\begin{equation}
 g_1 > 0, \ g_2 > 0, \ h < 0,  h + g' < 0, g_3 + g' + h > - \sqrt{g_1g_2}.
 \end{equation}
 The Higgs mass matrix can be simply diagonarized in the extreme cases, $\tan \beta \gg 1$, $\tan \beta \simeq 1$ and $\tan \beta   \ll 1$ \cite{HKMT}. In the first case; $\tan \beta \gg 1$, we get the relations between the parameters and mass eigenvalues,
\begin{equation}
g_2 = \frac{m_1^2}{2v^2}, \ g' + {\bar \xi} = \frac{m_2^2 + m_3^2}{2v^2}, \ h = \frac{m_2^2 - m_3^2}{2v^2}, {\bar \xi} = \frac{m_H^2}{v^2},\\ g' = \frac{m_2^2 + m_3^2 - 2m_H^2}{2v^2},
\end{equation}
and mass eigenvectors as
\begin{eqnarray}
u^{(1)} &=& \lbrack \cos \beta - \epsilon \sin \beta, \ - \sin \beta,\  0 \rbrack , \nonumber \\
u^{(2)} &=& \lbrack \sin \beta \cos \phi, \ (\cos \beta - \epsilon \sin \beta )\cos \phi,\  - \sin \phi \rbrack , \nonumber \\
u^{(3)} &=& \lbrack \sin \beta \sin \phi,\  (\cos \beta - \epsilon \sin \beta )\sin \phi,\ \cos \phi \rbrack ,
\end{eqnarray}
where $m_H$ is the mass of the charged Higgs boson, ${\bar \xi} = \xi/{\vert v_1v_2 \vert}^2$ and $$\epsilon \simeq \frac{2(2m_H^2 - m_1^2 -2g_3v^2)}{m_2^2 + m_3^2 -2m_1^2} \cos \beta.$$
In the next case; $\tan \beta \ll 1$, except for replacing $g_2$ with $g_1$ and $\cos \beta$ with $-\sin \beta$, the mass matrix is the same as in the case of $\tan \beta \gg 1$, so the $g_2$ is replaced by $g_1$ alone and the others remain the same as in Eq. (16). The eigenvectors are easily obtained as follows,
\begin{eqnarray}
u^{(1)} &=& \lbrack \cos \beta,\  - (\sin \beta + \epsilon' \cos \beta ), \  0 \rbrack ,\nonumber \\
u^{(2)} &=& \lbrack (\epsilon' \cos \beta  + \sin \beta )\cos \phi, \ \cos \beta \cos \phi, \ - \sin \phi \rbrack ,\nonumber \\
u^{(3)} &=& \lbrack ( \epsilon' \cos \beta + \sin \beta )\sin \phi,\ \cos \beta \sin \phi, \ \cos \phi \rbrack ,
\end{eqnarray} 
where $$\epsilon' \simeq - \lbrack 2(2m_H^2 - m_1^2 - 2g_3v^2)/(m_2^2 + m_3^2 -2m_1^2) \rbrack \sin \beta$$.

The last case to be considered is of $\tan \beta \simeq 1$. We get the same-type relations as follows,
\begin{equation}
h = \frac{m_2^2 - m_1^2 -m_3^2 + 2m_H^2}{2v^2} - g_3, \quad g' = \frac{m_2^2 + m_1^2 + m_3^2}{2v^2} - g_3,
\end {equation}
and eigenvectors as 
\begin{eqnarray}
u^{(1)}&=&(\cos \beta \cos \theta_{12} - \sin \beta \sin \theta_{12}\cos \theta_{23}, -\sin \beta \cos \theta_{12} - \cos \beta \sin \theta_{12}\cos \theta_{23}, \sin \theta_{12}\sin \theta_{23}), \nonumber \\
u^{(2)}&=&(\sin \beta \cos \theta_{12}\cos \theta_{23} + \cos \beta \sin \theta_{12}, \cos \beta \cos \theta_{12}\cos \theta_{23} - \sin \beta \sin \theta_{12}, -\cos \theta_{12} \sin \theta_{23}), \nonumber \\
u^{(3)}&=&(\sin \beta \sin \theta_{23}, \ \cos \beta \sin \theta_{23}, \ \cos \theta_{23}),
\end{eqnarray} 
where 
\begin{eqnarray}
\tan 2\theta_{12} &\simeq &(g_2 - g_1)/(m_1^2 - m_2^2), \nonumber \\
\tan 2\theta_{23} &\simeq &\lbrack (m_2^2 - m_1^2 - m_3^2 + 2m_H^2 -2g_3v^2)/(m_2^2 - m_3^2) \rbrack \tan 2\phi . \nonumber
\end{eqnarray} 

\section{The asymmetry $A^t$ in two-Higgs doublet model}

As stated in the last section, we calculate the asymmetry $A^t$ in the simplest extension of the Standard Model, that is, in the two-Higgs doublet model on account that this asymmetry is sensitive to the non-standard origin of CP violation. We estimate the asymmetry $A^t$ in Eq. (7).  The CP-violating part of the form factor $f_2^R$ can be obtained from the five one-loop diagrams of Fig. 1, since we use the unitary guage. We obtain the CP-violating contributions to the form factor ${\rm Im} f_2^R$ for the five diagrams as follows,
\begin{eqnarray}
{\rm Im} f_2^R|_1 &=& \frac{1}{(4\pi)^2} \frac{gG_F}{v}m_b^2m_t\sum_{k = 1}^3\lbrack{\vert v_1\vert}^2(-{\rm Im} Z_1^{(k)} + {\rm Im}{\tilde Z}_0^{(k)} ) (C_{12} + C_{23}) \nonumber \\
&& \mbox{}  - {\vert v_1\vert}^2 {\rm Im} Z_0^{(k)} (C_0 + C_{11} ) + {\vert v_2\vert}^2 {\rm Im} Z_0^{(k)} (C_{11} + C_{21} - C_{12} - C_{23})\rbrack, 
\end{eqnarray}
where $C...= C...(-p_t, p_{W^+}, m_b, m_H, m_k)$, 
\begin{eqnarray}
{\rm Im} f_2^R|_2 &=& -\frac{1}{(4\pi)^2}\frac{gG_F}{v}m_t\sum_{k=1}^3\lbrack m_b^2{\vert v_2\vert}^2(- {\rm Im} Z_2^{(k)} + {\rm Im} {\tilde Z}_0^{(k)})(C_{12} + C_{23} -C_{11} - C_{21}) \nonumber \\
&& \mbox{} + m_t^2{\vert v_1\vert}^2 \{ ({\rm Im} {\tilde Z}_0^{(k)} - {\rm Im} Z_2^{(k)})(C_0 + C_{11})  + {\rm Im} Z_0^{(k)} (C_{12} + C_{23})\} \rbrack, 
\end{eqnarray}
where $C... = C... (-p_b,  -p_{W^+}, m_t, m_H, m_k)$, 
\begin{eqnarray} 
{\rm Im} f_2^R|_3 &=& -\frac{1}{(4\pi)^2}\frac{gG_F}{v}m_b^2m_t{\vert v_2\vert}^2\sum_{k = 1}^3 {\rm Im} Z_0^{(k)}(C_0 + 2C_{11} + C_{21}), 
\end{eqnarray}
where $C... = C... (-p_t,  p_{W^+}, m_b, m_W, m_k)$, 
\begin{eqnarray}
{\rm Im} f_2^R|_4  &=& \frac{1}{(4\pi)^2} \frac{gG_F}{v}m_t{\vert v_1\vert}^2\sum_{k = 1}^3 {\rm Im }Z_0^{(k)}\lbrack m_b^2C_{31} + (m_t^2 - m_b^2 - m_W^2)C_{33} \nonumber \\
&& \mbox{} + m_W^2C_{34} + \frac16 - 6C_{35} + 2m_b^2C_{21} + m_W^2C_{22} + (m_t^2 -2m_b^2 - m_W^2)C_{23} \nonumber \\
&& \mbox{} - 4C_{24} + (m_b^2 - 2m_W^2)(C_{11} - C_{12}) + m_t^2(C_{12} + C_{23})\rbrack,
\end{eqnarray}
where $C... = C...(-p_b, -p_{W^+}, m_t, m_W, m_k)$, 
\begin{eqnarray}    
{\rm Im} f_2^R|_5 &=& \frac{1}{(4\pi)^2} \frac{gG_F}{v}v^2 m_b^2m_t\sum_{k = 1}^3\lbrack {\rm Im} Z_0^{(k)}(C_{11} - C_{12}) \nonumber \\
&& \mbox{} + {\rm Im} {\tilde Z}_0^{(k)} (C_{11} - C_{12} + C_{21} - C_{23} )\rbrack,
\end{eqnarray}
where $C... = C...(-p_b, -p_{W^+}, m_k, m_b, m_t)$. 

$C$... are the three-point functions of the loop integrals for the diagrams in Fig. 1 \cite{HHW}\cite{PV}. Now we evaluate the asymmetry $A^t$ in Eq. (7). The free parameters are $g_3$, $\phi$, $\tan \beta (= \vert v_2\vert/\vert v_1\vert )$, $m_1$, $m_2$, $m_3$ and $m_H$, and the Higgs masses are constrained by the positivity conditions of Eq. (15) to a certain extent. In our calculation, we will fix the following quantities as $g_3 \sim O(1)$, $\phi = \pi/6$, $m_H = 300$ GeV, $m_1 = 350$ GeV and $m_t = 180$ GeV. For the lightest Higgs boson($m_2$), we take $m_2 = 100$ GeV and 130 GeV in accord with the minimal supersymmetric standard model, or $m_2 = 160$ GeV as a numerical example of the eigenvalue. Because of the positivity condition, we obtain $m_3 > m_2$ and we adopt $110 \leq m_3 \leq 300$ GeV as the reasonable $m_3$ mass range. In the following calculation, we assume $|P_{||}^t| = 1$ in order to estimate the maximum value of $A^t$. 

The numerical results are shown in Figs. 2 $-$ 5. In Figs. 2 and 3, we show the $\tan \beta$-dependence of $A^t$ for $m_2 = 160$ GeV and $m_3 = 220$ GeV, and $m_2 = 130$ GeV and $m_3 = 190$ GeV, respectively. The results are obtained by using the solutions of mass eigenvalues $m_k \ (k = 1, 2, 3)$ and eigenvectors $u^{(k)}$ of Higgs mass matrix for the extreme cases, $\tan \beta \ll 1$, $\tan \beta \simeq 1$ and $\tan \beta \gg 1$, obtained in the last section. In Fig. 4, we show the $m_t$-dependence for $m_2 = 100$ GeV and $m_3 = 250$ GeV  in the range of $300 \geq m_t \geq 100$ GeV. In Fig. 5, we show the $m_3$-dependence for $m_2 = 100$ GeV in the range of $300 \geq m_3 \geq 110$ GeV. In Figs. 4 and 5, we have taken $\tan \beta = 0.5$.

As seen in Figs. 2 and 3, the values of $A^t$ are comparably smoothly connected among the three regions of $\tan \beta$, though the solutions for the mass and eigenvector of Higgs mass matrix are obtained in the extreme region of $\tan \beta$. The $\tan \beta$-dependence is nearly one order lower than the previous solutions \cite{HHW}, where only one generate neutral Higgs boson is adopted as an approximation. The reason why the magnitude of $A^t$ is reduced in the new calculation is that some cancellation has occurred in the evaluation of Im $f_2^R$ of Eqs. (21) - (25), since there are the relations of $\sum_{k = 1}^{3} ({\rm Im} Z_0^{(k)}, \ {\rm Im} {\tilde Z}_0^{(k)},\  {\rm Im} Z_1^{(k)}, \ {\rm Im} Z_2^{(k)}) = 0$. In the case of $\tan \beta \ll 1$, we get $1 \times 10^{-4} \sim 4 \times 10^{-3}$ for $A^t$. As seen in Fig. 4, the $m_t$-dependence is strong, as expected from its dependence of Higgs scalar coupling to the fermions.

\section{Conclusions and discussions}

We have investigated CP violation in the top-quark decay by studying the asymmetry $A^t$ of lepton distributions from the subsequent decay $W^+ \rightarrow l^+\nu$ in $t \rightarrow W^+b$ in the two-Higgs doublet model in three cases, $\tan \beta \ll 1$, $\tan \beta \gg 1$ and $\tan \beta \simeq 1$. As is expected, due to the large coupling of Higgs scalars to the top quark, the asymmetry is significantly large ($1 \times 10^{-4} \sim 4 \times 10^{-3}$) for the one-loop effects in the case of $\tan \beta \ll 1$, for typical parameter values of $m_H = 300$ GeV, $m_1 = 350$ GeV, $m_2 =  130 -160$ GeV, and $m_3 = 190 - 220$ GeV. The asymmetry turns out to be affected by $\tan \beta$ strongly in comparison with by $m_t$ and $m_3$. Hereafter, we are anxious to know about the value of $\tan \beta$.

\newpage
{\large\bf Figure Captions}
\vspace{ 0.5 cm}

Fig. 1. The five one-loop diagrams for the CP-violating part in $t \rightarrow W^+b$ decay in the two-Higgs doublet model in the unitary guage. The propagators with ${\phi}^0$ are the one of $\langle \phi_i^0 \phi_j^{0*}\rangle$ defined in Eq. (10).
\vspace{ 0.5 cm}

Fig. 2. The $\tan \beta$-dependence of $A^t$. (a) $\tan \beta \ll 1$, (b) $\tan \beta \simeq 1$, (c) $\tan \beta \gg 1$ for $m_H = 300$ GeV, $m_1 = 350$ GeV, $m_2 = 160$ GeV and $m_3 = 220$ GeV.
\vspace{ 0.5 cm}

Fig. 3. The $\tan \beta$-dependence of $A^t$. (a) $\tan \beta \ll 1$, (b) $\tan \beta \simeq 1$, (c) $\tan \beta \gg 1$ for $m_H = 300$ GeV, $m_1 = 350$ GeV, $m_2 = 130$ GeV and $m_3 = 190$ GeV.
\vspace{ 0.5 cm}

Fig. 4. The $m_t$-dependence of $A^t$ for $m_H = 300$ GeV, $m_1 = 350$ GeV, $m_2 = 100$ GeV, $m_3 = 250$ GeV and $\tan \beta = 0.5$.
\vspace{ 0.5 cm}
      
Fig. 5. The $m_3$-dependence of $A^t$ for $m_H = 300$ GeV, $m_1 = 350$ GeV, $m_2 = 100$ GeV, $m_t = 180$ GeV and $\tan \beta = 0.5$.

\newpage

\begin{figure}
\begin{center}
{\unitlength=1mm
\begin{picture}(120,210)
\thicklines
\put(9,32){\makebox(0,0){\large t}}
\put(12,32){\vector(1,0){6}}
\put(18,32){\line(1,0){6}}
\put(24,32){\vector(1,-1){7.12}}
\put(31.12,24.88){\line(1,-1){5.12}}
\multiput(24,32)(4.32,4.32){3}{\line(1,1){3.6}}
\put(36.24,19.76){\vector(0,1){13.24}}
\put(36.24,33){\line(0,1){11.24}}
\multiput(36.24,19.76)(2.88,-4.32){3}{\line(-1,-5){0.72}}
\multiput(39.12,15.44)(2.88,-4.32){3}{\line(-5,1){3.6}}
\put(36.24,44.24){\vector(1,1){8.64}}
\put(26.12,21.88){\makebox(0,0){\large t}}
\put(26.12,42.12){\makebox(0,0){\large $\phi^0$}}
\put(47.88,2.8){\makebox(0,0){\large ${\rm W}^+$}}
\put(47.88,54.88){\makebox(0,0){\large b}}
\put(42.24,32){\makebox(0,0){\large b}}
\put(9,54.88){\makebox(0,0){\large (5)}}

\put(9,98){\makebox(0,0){\large t}}
\put(12,98){\vector(1,0){6}}
\put(18,98){\line(1,0){6}}
\multiput(24,98)(2.88,-4.32){7}{\line(5,-1){3.6}}
\multiput(26.88,93.68)(2.88,-4.32){6}{\line(1,5){0.72}}
\put(24,98){\vector(1,1){7.12}}
\put(31.12,105.12){\line(1,1){5.12}}
\multiput(36.24,84.32)(0,5.48){5}{\line(0,1){4.0}}
\put(36.24,110.24){\vector(1,1){8.64}}
\put(25.12,87.88){\makebox(0,0){\large ${\rm W}^+$}}
\put(27.12,108.12){\makebox(0,0){\large b}}
\put(42.24,98){\makebox(0,0){\large $\phi^0$}}
\put(47.88,67.36){\makebox(0,0){\large ${\rm W}^+$}}
\put(47.88,120.88){\makebox(0,0){\large b}}
\put(9,120.88){\makebox(0,0){\large (3)}}

\put(64,98){\makebox(0,0){\large t}}
\put(67,98){\vector(1,0){6}}
\put(73,98){\line(1,0){6}}
\multiput(79,98)(4.32,-4.32){3}{\line(1,-1){3.6}}
\put(79,98){\vector(1,1){7.12}}
\put(86.12,105.12){\line(1,1){5.12}}
\multiput(91.24,110.24)(0,-4.08){6}{\line(2,-1){4.08}}
\multiput(91.24,106.16)(0,-4.08){6}{\line(2,1){4.08}}
\put(91.24,110.24){\vector(1,1){8.64}}
\multiput(91.24,85.76)(2.88,-4.32){3}{\line(-1,-5){0.72}}
\multiput(94.12,81.44)(2.88,-4.32){3}{\line(-5,1){3.6}}
\put(80.12,87.88){\makebox(0,0){\large $\phi^0$}}
\put(82.12,108.12){\makebox(0,0){\large t}}
\put(102.24,98){\makebox(0,0){\large ${\rm W}^+$}}
\put(102.88,67.36){\makebox(0,0){\large ${\rm W}^+$}}
\put(102.88,120.88){\makebox(0,0){\large b}}
\put(64,120.88){\makebox(0,0){\large (4)}}

\put(9,164){\makebox(0,0){\large t}}
\put(12,164){\vector(1,0){6}}
\put(18,164){\line(1,0){6}}
\multiput(24,164)(4.32,-4.32){3}{\line(1,-1){3.6}}
\put(24,164){\vector(1,1){7.12}}
\put(31.12,171.12){\line(1,1){5.12}}
\multiput(36.24,151.76)(0,5.12){5}{\line(0,1){4.0}}
\put(36.24,176.24){\vector(1,1){8.64}}
\multiput(36.24,151.76)(2.88,-4.32){3}{\line(-1,-5){0.72}}
\multiput(39.12,147.44)(2.88,-4.32){3}{\line(-5,1){3.6}}
\put(26.12,154.16){\makebox(0,0){\large ${\rm H}^+$}}
\put(27.12,174.12){\makebox(0,0){\large b}}
\put(42.24,164){\makebox(0,0){\large $\phi^0$}}
\put(47.88,133.36){\makebox(0,0){\large ${\rm W}^+$}}
\put(47.88,186.88){\makebox(0,0){\large b}}
\put(9,186.88){\makebox(0,0){\large (1)}}

\put(64,164){\makebox(0,0){\large t}}
\put(67,164){\vector(1,0){6}}
\put(73,164){\line(1,0){6}}
\multiput(79,164)(4.32,-4.32){3}{\line(1,-1){3.6}}
\put(79,164){\vector(1,1){7.12}}
\put(86.12,171.12){\line(1,1){5.12}}
\multiput(91.24,151.76)(0,5.12){5}{\line(0,1){4.0}}
\put(91.24,176.24){\vector(1,1){8.64}}
\multiput(91.24,151.76)(2.88,-4.32){3}{\line(-1,-5){0.72}}
\multiput(94.12,147.44)(2.88,-4.32){3}{\line(-5,1){3.6}}
\put(80.12,154.16){\makebox(0,0){\large $\phi^0$}}
\put(82.12,174.12){\makebox(0,0){\large t}}
\put(97.24,164){\makebox(0,0){\large ${\rm H}^+$}}
\put(102.88,133.36){\makebox(0,0){\large ${\rm W}^+$}}
\put(102.88,186.88){\makebox(0,0){\large b}}
\put(64,186.88){\makebox(0,0){\large (2)}}

\end{picture}}
\end{center}
\caption{ }
\end{figure}

\newpage
\begin{figure}[t]
\epsfile{file=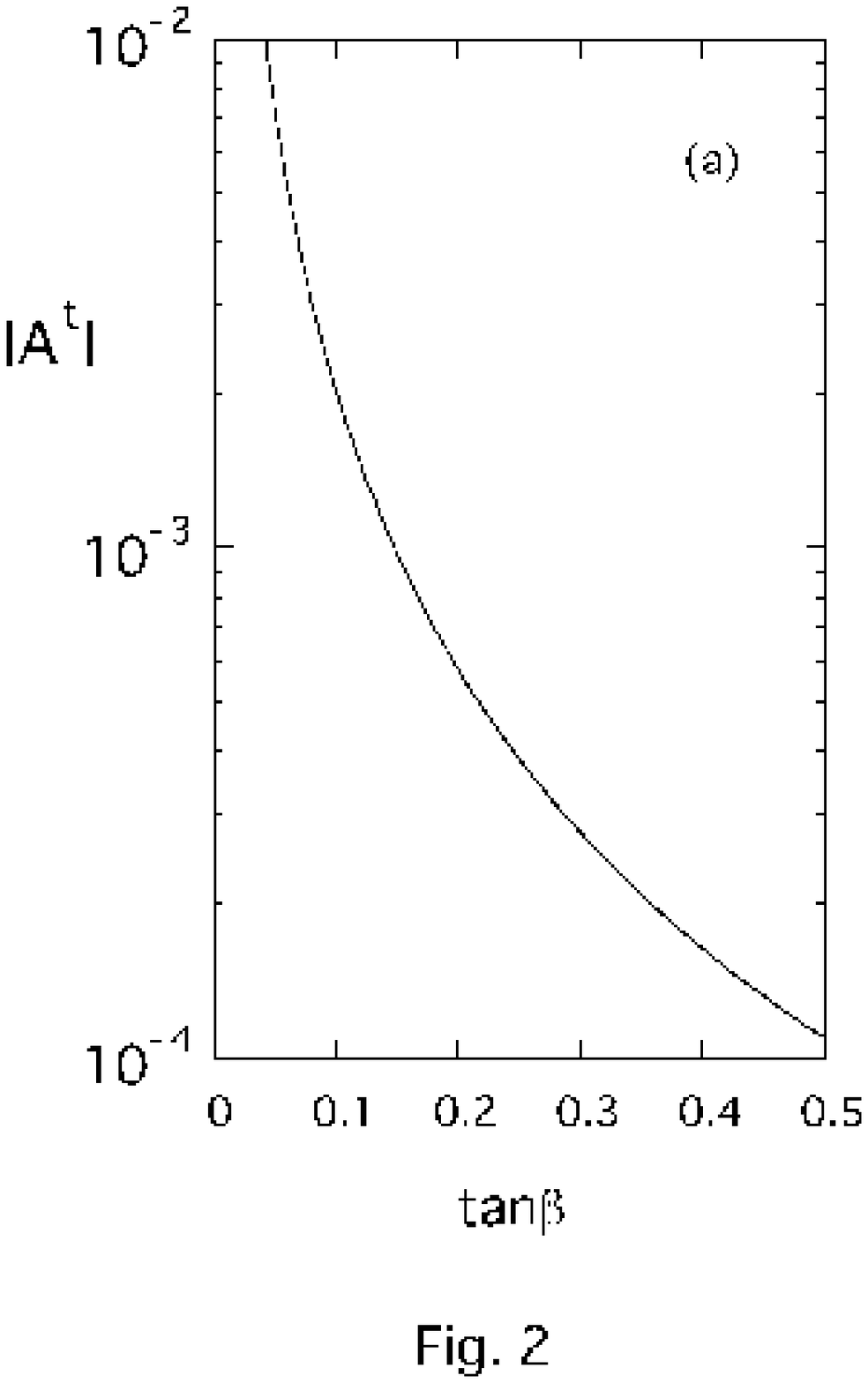,width=15cm}
\end{figure}

\newpage
\begin{figure}[t]
\epsfile{file=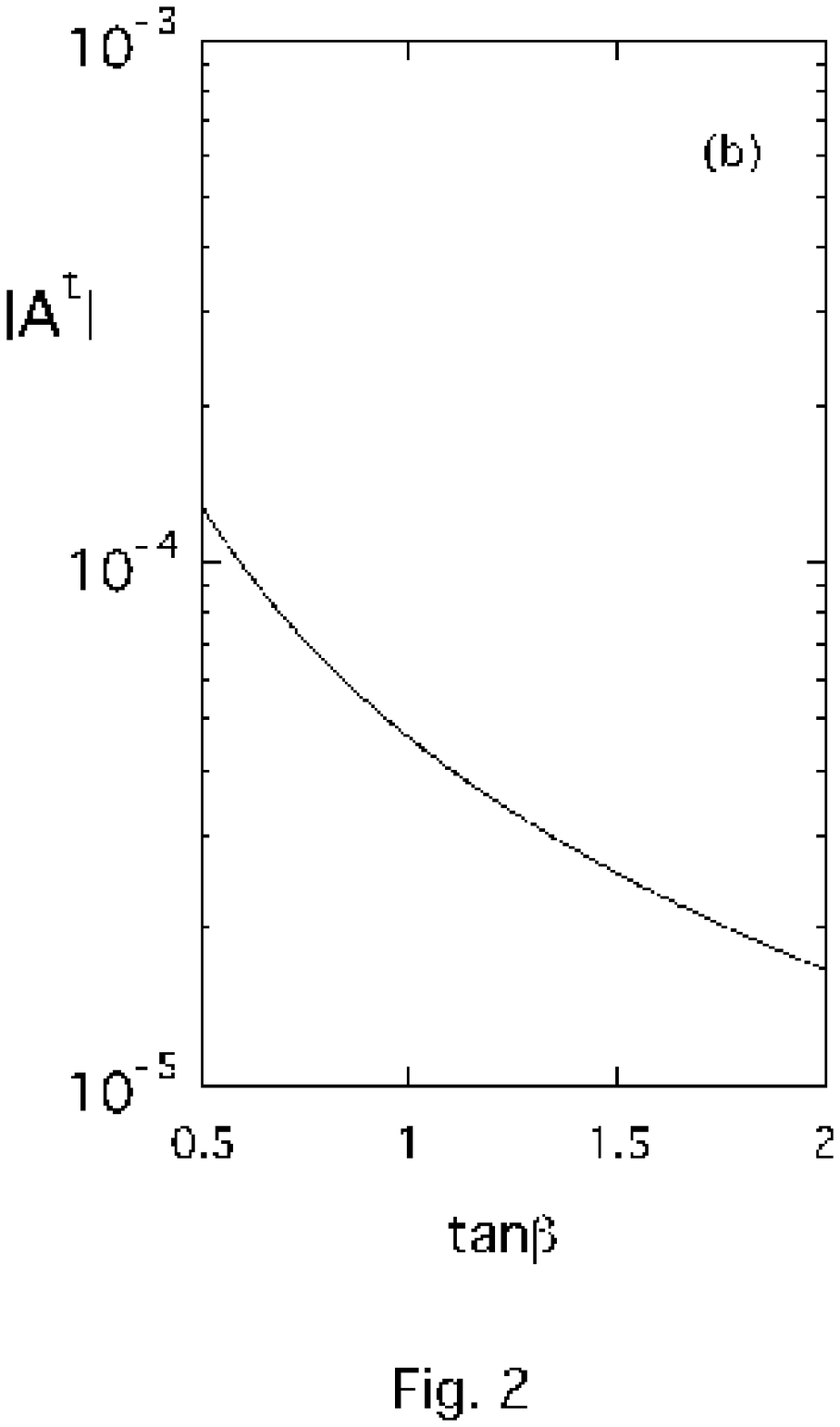,width=15cm}
\end{figure}

\newpage
\begin{figure}[t]
\epsfile{file=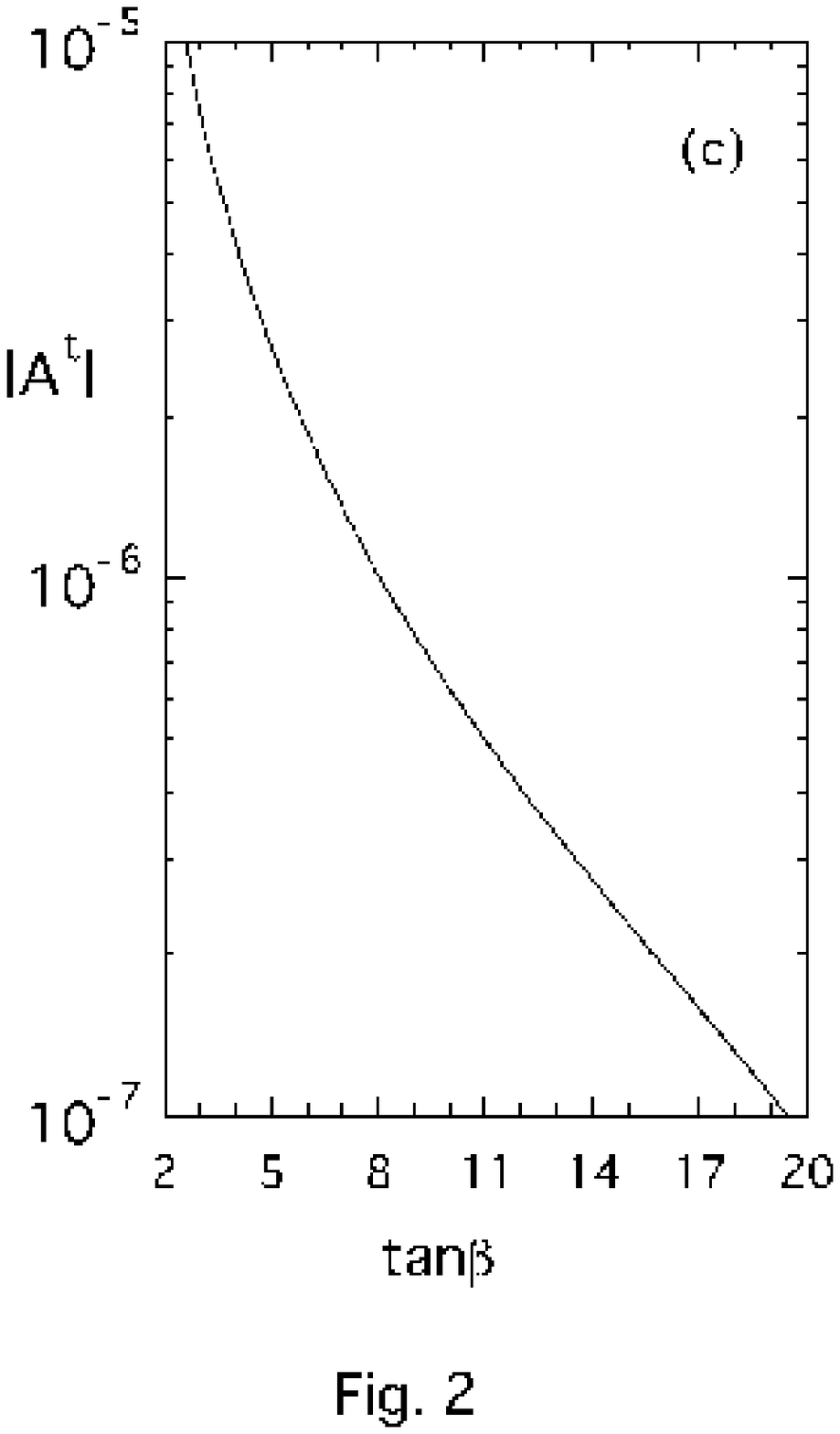,width=15cm}
\end{figure}

\newpage
\begin{figure}[t]
\epsfile{file=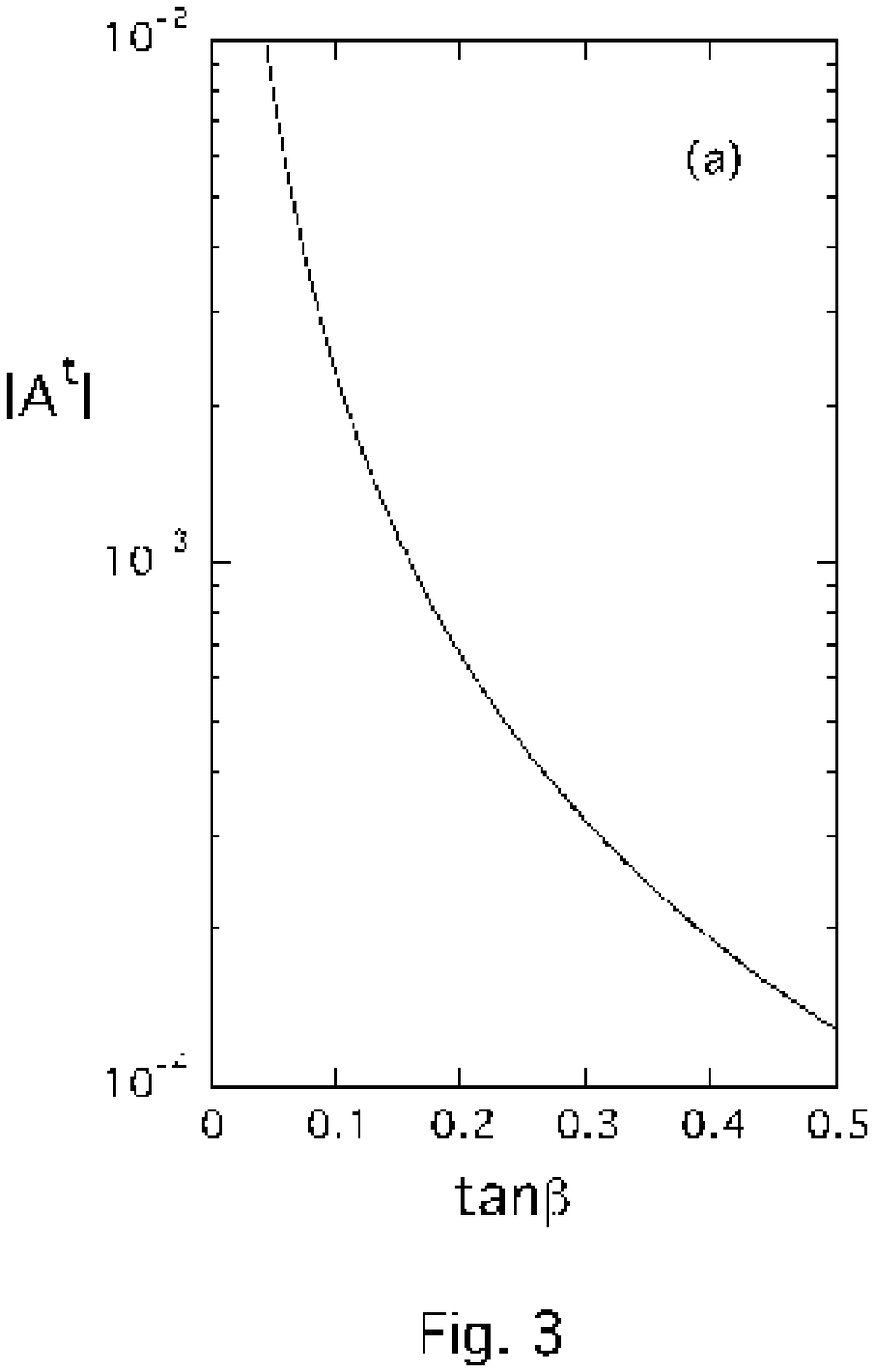,width=15cm}
\end{figure}

\newpage
\begin{figure}[t]
\epsfile{file=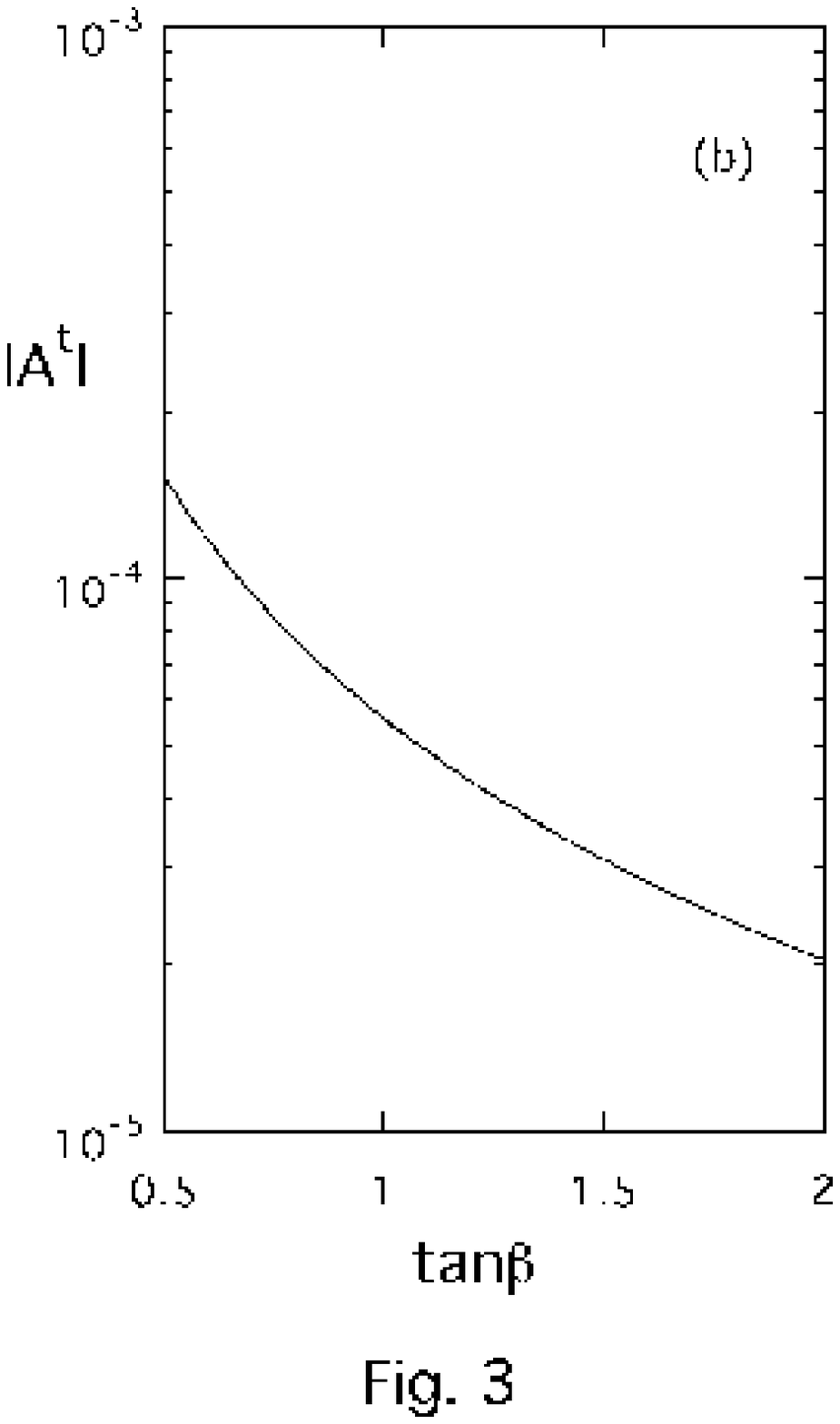,width=15cm}
\end{figure}

\newpage
\begin{figure}[t]
\epsfile{file=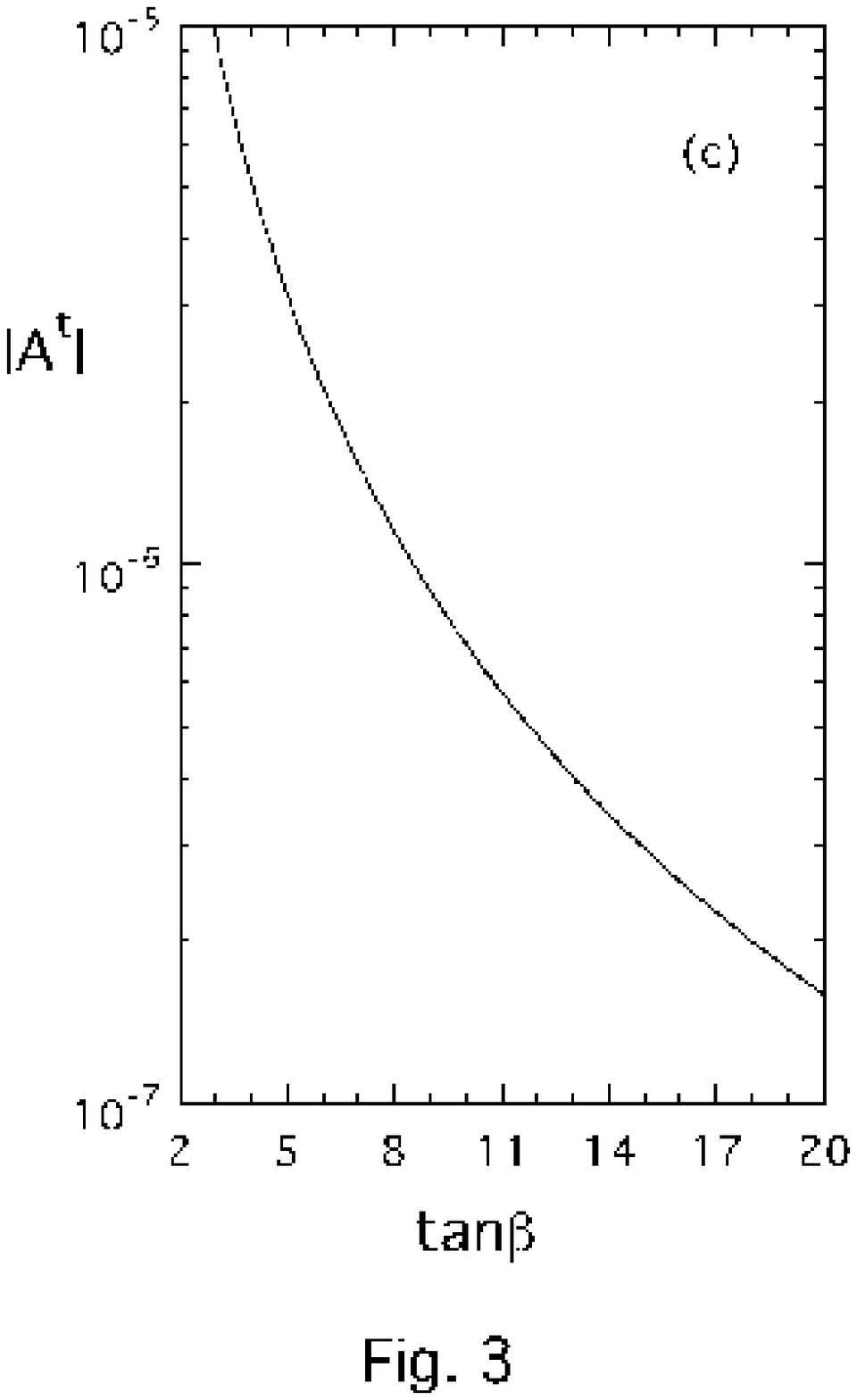,width=15cm}
\end{figure}

\newpage
\begin{figure}[t]
\epsfile{file=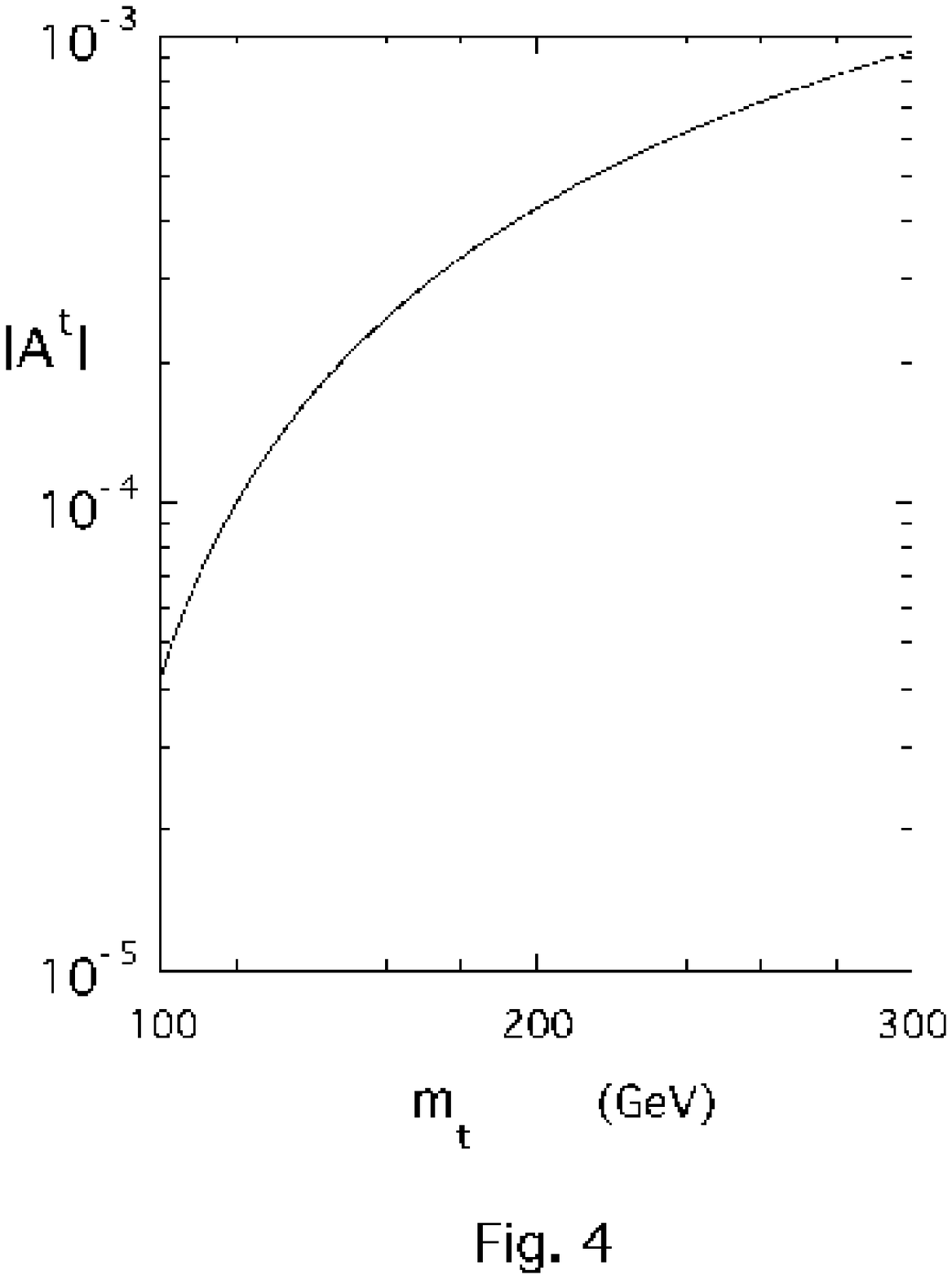,width=16cm}
\end{figure}

\newpage
\begin{figure}[t]
\epsfile{file=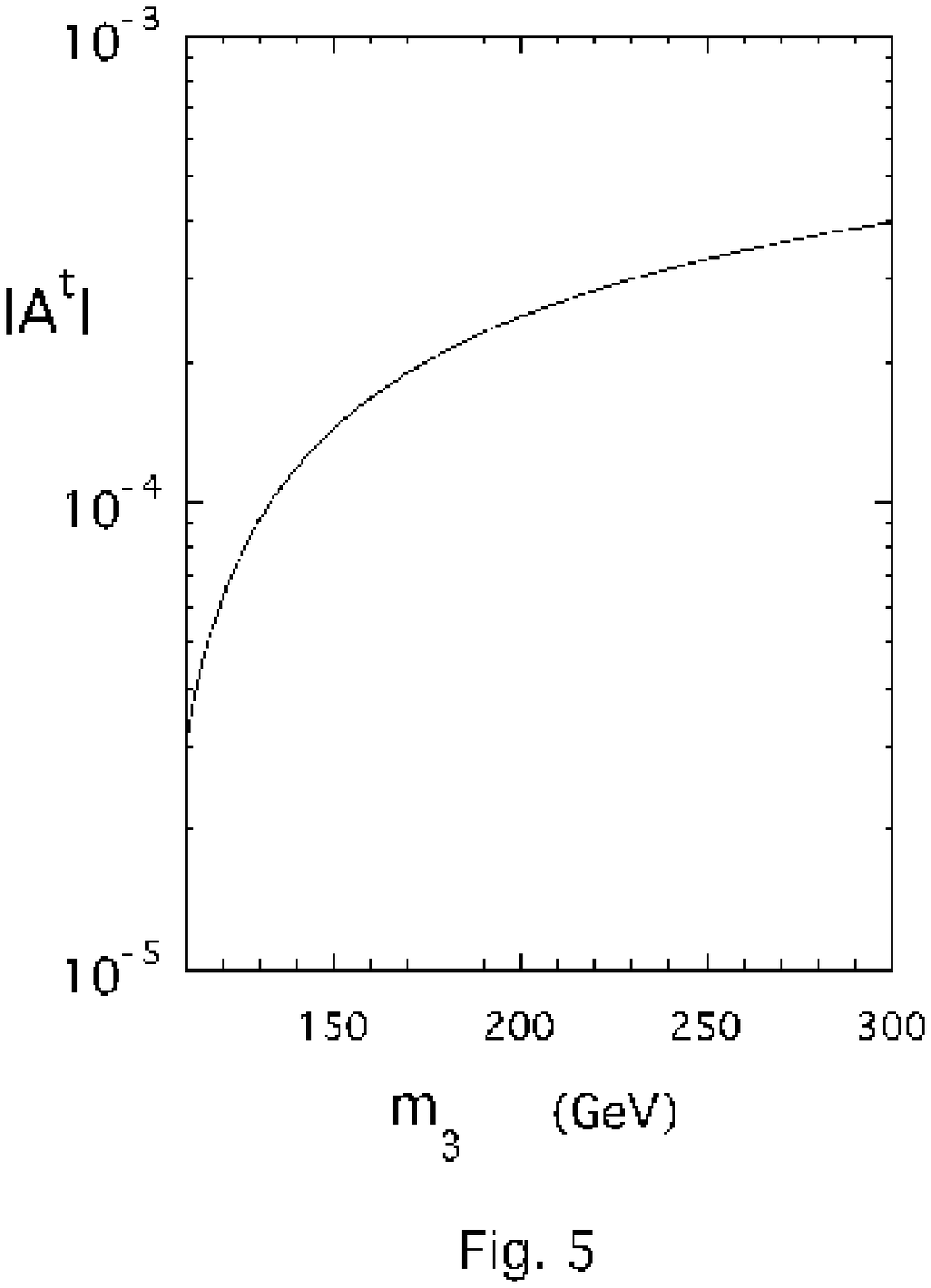,width=16cm}
\end{figure}

\end{document}